\documentclass[aps,prb,twocolumn,superscriptaddress,amsmath]{revtex4-2}
\usepackage[colorlinks=true, linkcolor=blue, filecolor=blue, urlcolor=blue, citecolor=blue]{hyperref}
\usepackage{amsfonts}
\usepackage{bbm}
\usepackage{graphicx}
\usepackage{dcolumn}
\usepackage{bm}
\usepackage{soul}

\begin{document}

\title{Universal theory of tunable second-order topological corner states induced by interlayer coupling in twist bilayer Chern insulators}

\author{Cheng-Ming Miao}
\email[]{These authors contribute equally to this work.}
\affiliation{International Center for Quantum Materials, School of Physics, Peking University, Beijing 100871, China}

\author{Yu-Hao Wan}
\email[]{These authors contribute equally to this work.}
\affiliation{International Center for Quantum Materials, School of Physics, Peking University, Beijing 100871, China}

\author{Ying-Tao Zhang}
\affiliation{College of Physics, Hebei Normal University, Shijiazhuang 050024, China}

\author{Qing-Feng Sun}
\email[]{sunqf@pku.edu.cn}
\affiliation{International Center for Quantum Materials, School of Physics, Peking University, Beijing 100871, China}
\affiliation{Hefei National Laboratory, Hefei 230088, China}

\begin{abstract}
We propose a universal theory for tunable second-order topological corner states induced by interlayer coupling in bilayer Chern insulators with opposite Chern numbers.
We demonstrate that the existence of the topological corner state is determined by the relationship between the twist angle of the bilayer Chern insulators and the normal angles of the two sides of the corner.
In addition, the position of these corner states can be sensitively controlled by the twist angle, as confirmed by a rigorous analysis of edge state theory.
Our findings serve as a universal theory, opening avenues for the design and realization of higher-order topological materials.
\end{abstract}
\maketitle

\section{\label{sec1}Introduction}

Topological insulators are a class of materials characterized by the bulk-edge correspondence, i.e., insulating bulk properties and robust conducting states at the edges or surfaces \cite{Kane2005,Kane2005a,Bernevig2006,Qi2006,Jiang2009,Moore2010,Hasan2010,
Qi2011}.
In two-dimensional materials, time-reversal symmetry (TRS) plays a critical role in defining the nature of these edge states.
For instance, $\mathbb{Z}_2$ topological insulators, protected by TRS, are described by $\mathbb{Z}_2$ topological invariant and host helical edge states where the spin and momentum are locked \cite{Kane2005,Kane2005a,Bernevig2006,Kruthoff2017,Zhang2019,Vergniory2019,Tang2019}.
On the other hand, Chern insulators (CIs), which break TRS, are defined by a non-zero Chern number and exhibit chiral edge states that propagate unidirectionally \cite{Yu2010,Chang2013,Wang2013,Qiao2014,Sun2019,Chang2023,Mei2024}.
The Bernevig-Hughes-Zhang (BHZ) model \cite{Bernevig2006} is the simplest model of $\mathbb{Z}_2$ topological insulators, providing a theoretical framework for understanding helical edge states that are protected by TRS.
A half-BHZ model, also known as the Qi-Wu-Zhang model \cite{Qi2006}, serves as a fundamental model for exploring CIs in the absence of TRS.

Recent advances have introduced higher-order topological insulators, which exhibit additional localized states at corners or hinges beyond edge and surface states \cite{Benalcazar2017,Benalcazar2017a,Song2017,Langbehn2017,Peng2017,Ezawa2018,Ezawa2018a,Schindler2018,Kunst2018,Khalaf2018,Fukui2018,Banerjee2020,addr1,addr2,addr3,addr4}. These developments extend the bulk-edge correspondence of topological systems, summarizing that a $d$-dimensional $n$-th order topological system hosts gapless edge states in $d-n$ dimensions. Experimentally, second-order topological zero-dimensional corner states have been realized in various systems such as electrical circuits \cite{Imhof2018, Peterson2018, Ezawa2019, Serra-Garcia2019, Zhang2021}, acoustic \cite{Ni2018, Xue2018, Xue2019, Xue2020, Gao2021}, photonic crystals \cite{Li2019, Xie2019, Chen2019, Mittal2019, Hassan2019, Kim2020, Mandal2024}, and mechanical \cite{Wakao2020} $et~al$.
Although higher-order topological insulators in electronic systems have not yet been experimentally realized, several candidate systems and materials have been predicted to be higher-order topological insulators, such as breathing Kagome and pyrochlore lattices \cite{Ezawa2018}, black phosphorene \cite{Ezawa2018a}, graphdiyne \cite{Sheng2019}, twisted bilayer graphene \cite{Park2019, Spurrier2020}. In addition, there is a theoretical scheme to induce second-order topological corner states by introducing magnetism to break the TRS that protects helical edge states in first-order $\mathbb{Z}_2$ topological insulators \cite{Ren2020,Zeng2022,Han2022,Zhu2022,Miao2022,Miao2023,Zhu2023,Zhu2023a,Miao2024}.
The origin of these topological corner states can be easily understood in terms of the Dirac equation: the introduction of magnetism leads to the formation of Dirac mass domain walls at the corner, thereby binding the zero-dimensional corner states \cite{Jackiw1976}.
Naturally, the introduction of magnetism to induce higher-order topology is not applicable in first-order CIs.
Recent theoretical studies have demonstrated that higher-order topological phase transitions in CIs can be realized by coupling two CIs with opposite Chern numbers \cite{Mandal2024,Liu2024,Liu2024a}. However, the positions of the corner states are not tunable, and the underlying principle remains unclear.

In this paper, we propose a universal theory for second-order topological corner states induced by interlayer coupling in twist bilayer CIs.
As shown in Fig. \ref{fig1}(a), the top CI layer with $C = -1$ (in grey) can be obtained by reversing the bottom CI layer with $C = +1$ (in yellow) and twisting it by an angle $\alpha$.
The red (blue) edge state propagates counterclockwise (clockwise) in the top (bottom) layer.
Utilizing the Qi-Wu-Zhang model, we demonstrate that interlayer coupling
induces second-order topological corner states at angle $\theta=\alpha/2, \alpha/2+\pi$ [see Fig. \ref{fig1}(b)] where the edge states remain gapless,
while the edges at other angles are gapped by interlayer coupling.
So the positions of corner states can be controlled by the twist angle $\alpha$.
We use edge theory to derive the analytic conditions for the appearance of corner states.
Additionally, considering an angle consisting of the twist bilayer CIs,
if the normal angles ($\theta_1$, $\theta_2$) of the two sides of the angle satisfy the relation $\min(\theta_{1},\theta_2) <\alpha/2 (\alpha/2 +\pi)<\max(\theta_{1},\theta_2)$,
there is a zero-dimensional corner state at this angle.
This universal theory can explain all previous work on inducing topological corner states in two-dimensional systems with interlayer coupling.
Applying this theory, we design L-shaped coupled bilayer CIs with opposite Chern numbers, where the number of corner states can be tuned by twist angle.
This also demonstrates that the realization of corner states is not limited by nanoflake shape, providing a universal theory.

The rest of the paper is organized as follows.
In Sec. \ref{sec2}, we derive the tight-binding Hamiltonian for the coupled system composed of two CIs with opposite Chern numbers.
In Sec. \ref{sec3A}, we present the distribution of $\bf{d}$-vector in the first Brillouin zone and the in-gap states near zero energy to visually show the emergence and tunability of the second-order topological corner states.
In Sec. \ref{sec3B}, we analyze the conditions for the existence of the corner states.
We apply the universal theory to analyse the corner states in
the L-shaped twist bilayer CIs in Sec. \ref{sec3C}.
Finally, a summary is presented in Sec. \ref{sec4}.

\section{\label{sec2}model and hamiltonian}

\begin{figure}
	\centering
	\includegraphics[width=1\columnwidth,clip]{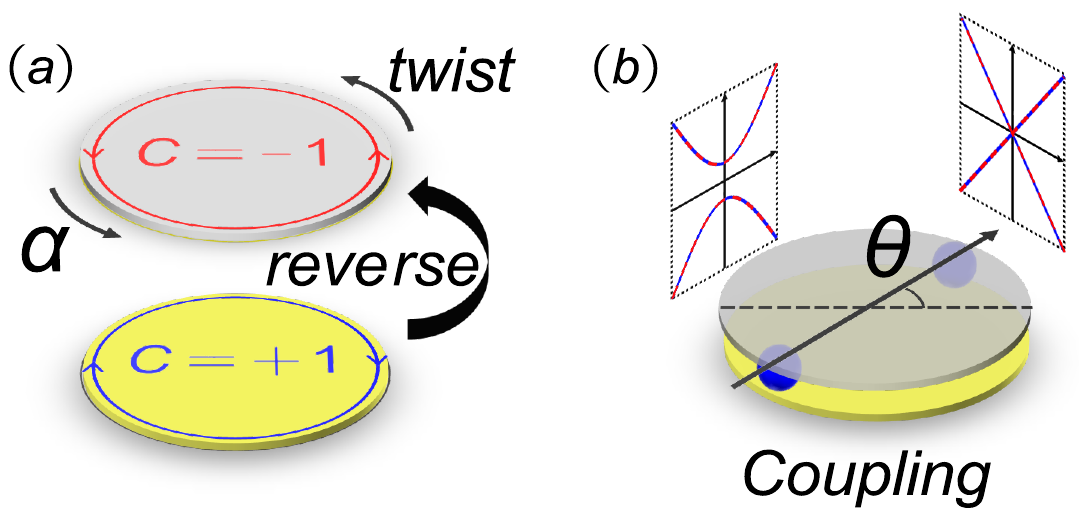}
	\caption{(a) Schematic plot of two decoupled CI layers with opposite Chern numbers, where the edge state propagates counterclockwise/clockwise in the top/bottom layer. The top layer $C = -1$ CI (in gray) can be obtained by inverting and twisting the bottom layer $C = +1$ CI (in yellow). $\alpha$ is the twist angle. (b) Coupling disrupts the two chiral edge states with opposite propagation directions, resulting in two corner states represented by blue spheres.
The two spectrum plots schematically show the $\theta$-dependent edge gap.
	}
	\label{fig1}
\end{figure}

We consider a coupled system composed of two circular nanoflake layers with opposite Chern numbers, as shown in Fig.~\ref{fig1}. The system Hamiltonian
in the momentum space for the two coupled CIs is given by:
\begin{align}
  H(\mathbf{k})=\left(\begin{array}{cc}
			H_T(\mathbf{k}) & t \sigma_{0}\\
			(t \sigma_{0})^*   & H_B(\mathbf{k})
		\end{array}\right),
  \label{eq1}
 \end{align}
where $\mathbf{k} = (k_x, k_y)$ is a wave vector in the first Brillouin zone,
$t$ represents the coupling strength between the top and bottom layers,
and $\sigma_{0}$ denotes the $2\times 2$ identity matrix.
$H_{T}(\mathbf{k})$ and $H_{B}(\mathbf{k})$ are the Hamiltonian for the top and bottom CI layers, respectively. We choose the Qi-Wu-Zhang model as an example of CIs \cite{Qi2006}, with the low-energy effective Hamiltonian $H_{T/B}(\mathbf{k})$ expressed as follows:
\begin{align}
  H_{T/B}(\mathbf{k})=\sum_{k}{\mathbf{d}_{T/B}(\mathbf{k}) \cdot \bm{\sigma}},
  \label{eq2}
 \end{align}
where $\mathbf{d}_{T/B}(\mathbf{k})$ is a vector with three components being given functions of $\mathbf{k}$ and $\bm{\sigma}=(\sigma_{x}, \sigma_{y}, \sigma_{z})$ are the Pauli matrices.
The representations for the components of the vector $\mathbf{d}_{T/B}(\mathbf{k})$ are given below:
  \begin{align}
  d_{T}^{x}(\mathbf{k})&=-A k_{x} \cos {\alpha}-A k_{y} \sin {\alpha}, \quad d_{B}^{x}(\mathbf{k})=Ak_{x},\nonumber \\
  d_{T}^{y}(\mathbf{k})&=-A k_{x} \sin {\alpha}+A k_{y} \cos {\alpha}, \quad d_{B}^{y}(\mathbf{k})=Ak_{y}, \nonumber \\
  d_{T}^{z}(\mathbf{k})&=M-B k^{2},\quad \quad \quad \quad \quad \quad \quad d_{B}^{z}(\mathbf{k})=M-B k^{2},
  \label{eq3}
 \end{align}
where $\alpha$ denotes the twist angle between the top and bottom layers.
$M$ is the Dirac mass term and $A$ and $B$ are the parameters of the model.
In the momentum space, the vector $\mathbf{d}_{T/B}(\mathbf{k})$ exhibits a Skyrmion configuration with winding number $n=\pm 1$ for $M/B> 0$. Since the tight-binding representation is used in our calculations, the above Hamiltonian can be mapped onto a tight-binding representation on a two-dimensional square lattice as:
  \begin{align}
  H=&\sum_{\bf i}^{}{[\psi_{\bf i}^{\dagger}T_{0}\psi_{\bf i}+(\psi_{\bf i}^{\dagger}T_{x}\psi_{{\bf i}+\delta\mathbf{x}}+\psi_{\bf i}^{\dagger}T_{y}\psi_{{\bf i}+\delta\mathbf{y}})+\mathrm{H.c.}]},\nonumber \\
  T_{0}=&(M-4B)\tau_{0}\sigma_{z}+t \tau_{x}\sigma_{0},\nonumber \\
  T_{x}=&B\tau_{0}\sigma_{z}+\frac{A}{4i}\tau_{0}[(1-\cos \alpha)\sigma_{x}-\sin \alpha \sigma_{y}]\nonumber \\&-\frac{A}{4i}\tau_{z}[(1+\cos \alpha)\sigma_{x}+\sin \alpha \sigma_{y}],\nonumber \\
  T_{y}=&B\tau_{0}\sigma_{z}+\frac{A}{4i}\tau_{0}[(1+\cos \alpha)\sigma_{y}-\sin \alpha \sigma_{x}]\nonumber \\&-\frac{A}{4i}\tau_{z}[(1-\cos \alpha)\sigma_{y}+\sin \alpha \sigma_{x}],
  \label{eq4}
 \end{align}
where $\psi_{\bf i}^{\dagger}=(c_{{\bf i},\uparrow, t}^{\dagger},c_{{\bf i},\downarrow, t}^{\dagger},c_{{\bf i},\uparrow, b}^{\dagger},c_{{\bf i},\downarrow, b}^{\dagger})$ and $c_{{\bf i}, \uparrow / \downarrow, t/b}^{\dagger}$ is the creation operator for an electron with pseudo-spin up/down ($\uparrow/\downarrow$) at the ${\bf i}$-th site of the top/bottom ($t/b$) layer.
${\bf i}=(x,y)$ is the coordinates of the size with $x$ and $y$ being integers
and $\delta \mathbf{x}$ $(\delta \mathbf{y})$ is the unit vector along the $x$ ($y$) direction.
$\tau_{0}$ and $\tau_{x,y,z}$ are the $ 2\times 2$ unit matrix and the Pauli matrices acting on layer degree of freedom.
The lattice constant has been set to 1 here.
Without loss of generality, we set $A=B=1$ and $M = 1$  in our calculations unless otherwise noted.

\section{\label{sec3}results and discussion}

\subsection{\label{sec3A}Tunable second-order topological corner states induced by interlayer coupling}

\begin{figure*}
	\centering
	\includegraphics[width=2.1\columnwidth,clip]{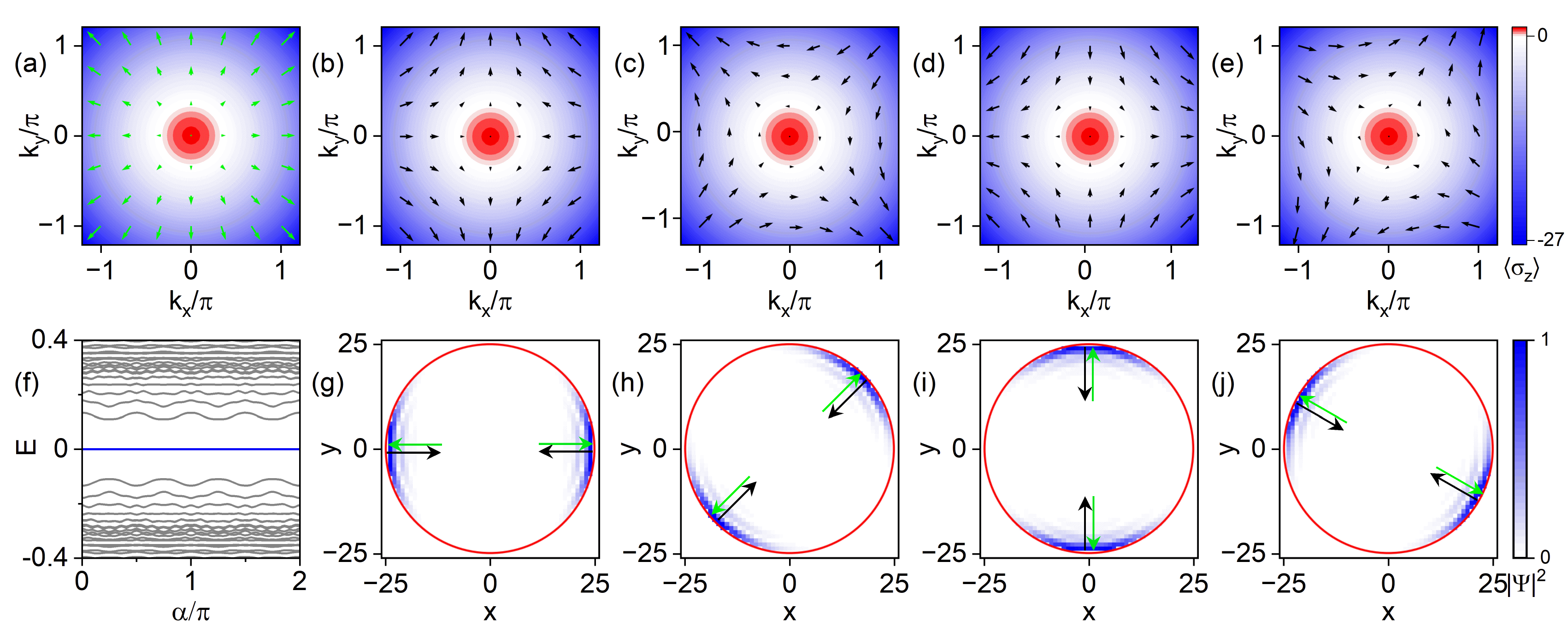}
	\caption{ (a-e) The distribution of $\mathbf{d}_{T/B}(\mathbf{k})$ in the first Brillouin zone for the system with $M=1$ and $A=B=1$. The color indicates the normal component of the plane, with red denoting the upward direction and blue the downward direction.
The green arrows in (a) indicate the direction of the in-plane spin components in the $C = +1$ CI, while the black arrows in (b-e) indicate the direction of the in-plane spin components in the $C = -1$ CI with different twist angles $\alpha$.
(f) Energy spectrum for the circular coupled bilayer CIs system as a function of the twist angle $\alpha$.
(g-j) Probability distribution of the in-gap states with different $\alpha$, where the brightness of blue is proportional to the square of the wave function [see color bar].
The red circular lines indicate the boundary of the circular nanoflake.
The green and black arrows indicate the directions of the in-plane spin components for monolayer CIs with Chern numbers $C = +1$ and $C = -1$, respectively. The twist angle are $\alpha = 0$ for (b,g), $\alpha = \pi/2$ for (c,h), $\alpha = \pi$ for (d,i), and $\alpha = 5\pi/3$ for (e,j). Other parameters are selected as coupling strength $t=0.3$ and circular nanoflake radius $R=25a$ for (f-g).
}
	\label{fig2}
\end{figure*}

In this subsection, we show the distribution of $\mathbf{d}_{T/B}(\mathbf{k})$ in the decoupled CI layers, described by the Hamiltonian in Eq.~(\ref{eq3}).
At the origin of the Brillouin zone $(k_x, k_y)=(0,0)$, the vector $\mathbf{d}_{T/B}(\mathbf{k})=(0,0,M)$. This implies that the
$\mathbf{d}_{T/B}(\mathbf{k})$ vector points along the $z$-axis at the origin, as illustrated in Figs. \ref{fig2}(a-e).
We focus on the behavior of $\mathbf{d}_{B}(\mathbf{k})$ in the two-dimensional $k_x\mbox{-}k_y$ plane, as represented by the green arrows in Fig. \ref{fig2}(a).
It can be seen from Fig. \ref{fig2}(a) that the winding number of $\mathbf{d}_{B}(\mathbf{k})$ is $n=+1$.
Furthermore, the black arrows in Figs. \ref{fig2}(b-e) depict the distribution of the $\mathbf{d}_{T}(\mathbf{k})$ vector in the plane with different twist angles $\alpha$.
While the distribution of the $\mathbf{d}_{T}(\mathbf{k})$ vector changes with $\alpha$, it consistently maintains a winding number of $n=-1$, as shown in Figs. \ref{fig2}(b-e).
This consistency is expected, as twisting a two-dimensional nanoflake by an angle $\alpha$ does not alter its topological properties.

Next, two circular nanoflakes with opposite Chern numbers are coupled to form a coupled bilayer system, as shown in Fig. \ref{fig1}(b).
To explore the second-order topology, we plot the energy spectrum for the coupled system versus twist angle $\alpha$ with an interlayer coupling strength $t=0.3$ in Fig. \ref{fig2}(f).
The geometry of the circular nanoflakes is set as: we construct a square lattice plane centered at the origin $(0,0)$, where the position of each site is labeled by its coordinates $(x,y)$. The sites belonging to the circular nanoflake satisfy the condition $x^{2}+y^{2} \leq R^{2}$ with radius $R=25$.
It is evident that the zero-energy in-gap states (highlighted in blue lines) remain stable throughout the variation of $\alpha$ [see Fig. \ref{fig2}(f)].
Moreover, we plot the distribution of the in-gap states with different twist angles $\alpha$ in Figs. \ref{fig2}(g-j).
The square of the wave function is scaled so that its maximum value is 1 (in blue), with the color transitioning to white as the value decreases to 0 [see color bar].
We illustrate the in-plane vectors $\mathbf{d}_{T}(\mathbf{k})$ and $\mathbf{d}_{B}(\mathbf{k})$ at the maximum positions of the state density distribution, represented by black and green arrows, respectively.
As shown in Fig. \ref{fig2}(g) with $\alpha=0$,
the in-gap states localize at the left and right corners of the circular nanoflake.
Notably, the directions of the vectors $\mathbf{d}_{T}(\mathbf{k})$ and $\mathbf{d}_{B}(\mathbf{k})$ remain exactly opposite at the positions of the corner states.
It can be seen from Fig. \ref{fig2}(h) that the zero-dimensional corner states are bound at the upper-right and lower-left corners of the circular nanoflake while $\alpha=\pi/2$.
As the $\alpha$ increases to $\pi$, the position of the corner states shift to the upper and lower corners in Fig. \ref{fig2}(i).
For the cases of $\alpha=\pi/2, \pi$, the vectors $\mathbf{d}_{T}(\mathbf{k})$ and $\mathbf{d}_{B}(\mathbf{k})$ remain always opposite at the corner states
[see Fig. \ref{fig2}(h,i)].
For a more general twist angle $\alpha=5\pi/3$, the corner state still occurs at the position where the vectors $\mathbf{d}_{T}(\mathbf{k})$ and $\mathbf{d}_{B}(\mathbf{k})$ are oppositely aligned [see Fig. \ref{fig2}(j)].
These results intuitively demonstrate that zero-dimensional corner states can be induced by interlayer coupling in bilayer CIs with opposite Chern numbers, and their positions can be finely tuned by varying the twist angle $\alpha$.

\subsection{\label{sec3B}The origin of second-order topological corner states}

 \begin{figure}
	\centering
	\includegraphics[width=1\columnwidth,clip]{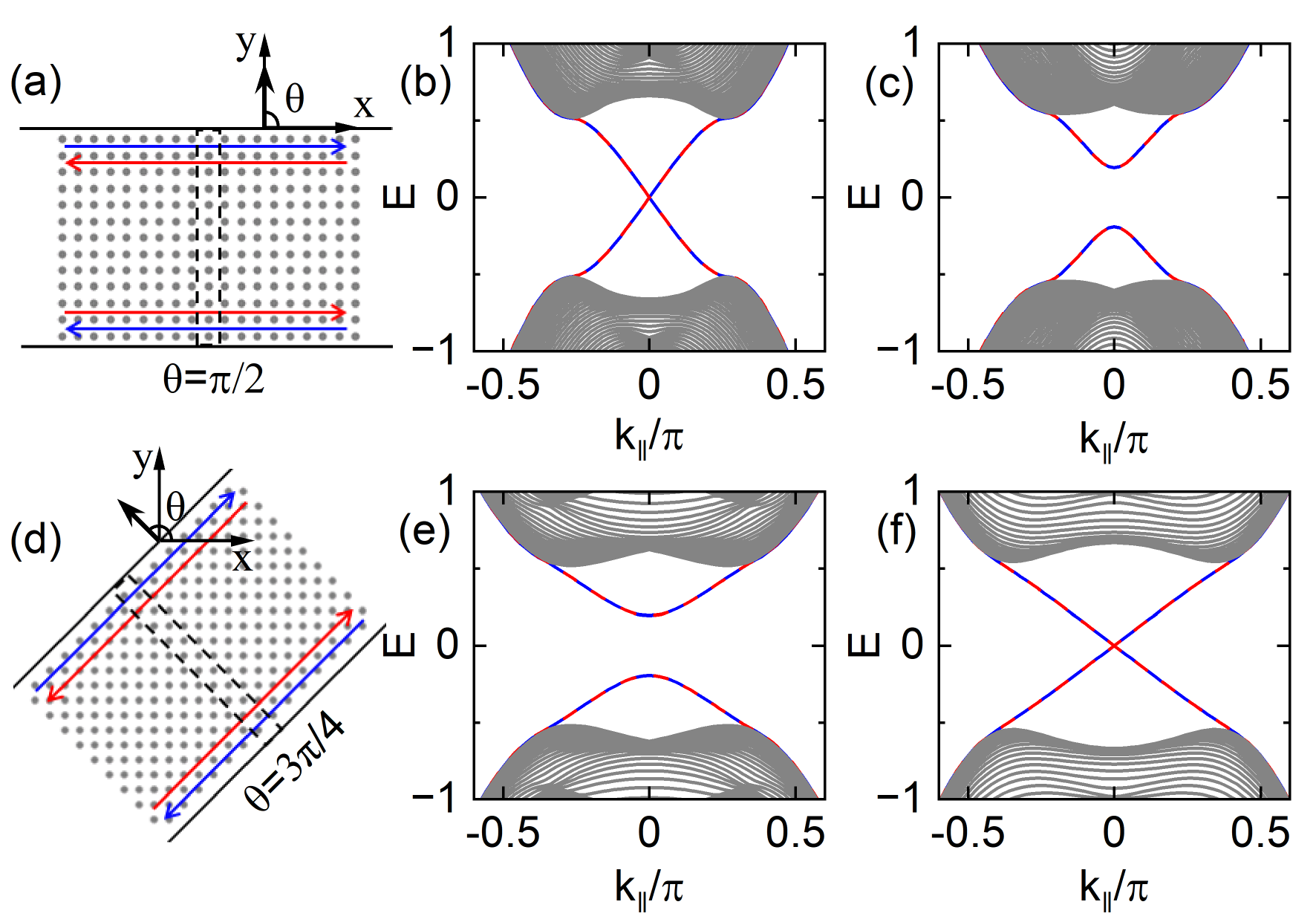}
	\caption{ Schematic diagram of nanoribbons with different normal angles of edges $\theta=\pi/2$ for (a) and $\theta=3\pi/4$ for (d). $\theta$ is the angle between the normal direction of the nanoribbon edges (black solid lines) and the $+x$ direction. Lattice sites are denoted by grey dots and the choice of super unit cells is illustrated by the black dashed box. Red and blue arrows indicate pseudospin-protected gapless edge states.
Panels (b,c) and (e,f) respectively display the band structures of nanoribbons corresponding to (a) and (d) with different twist angles.
The twist angles are $\alpha=\pi$ in (b,e) and $\alpha=3\pi/2$ in (c,f).
The width of the nanoribbon is $N=80$, and other parameters are the same as in Fig. \ref{fig2}(f).
	}
	\label{fig3}
\end{figure}

To verify that only tangential edges, where the corner states are located, maintain gapless energy bands, we examine the nanoribbons with different edge's normal angles $\theta=\pi/2, 3\pi/4$, as shown in Fig. \ref{fig3}(a,d).
$\theta$ is the angle between the normal direction of the nanoribbon edges
and the $+x$ direction. The nanoribbon width is chosen as $N = 80$.
We select the super unit cell (enclosed by black dashed box) in Fig. \ref{fig3}(a) to plot the energy band structure of the coupled twist bilayer CIs nanoribbon at edge $\theta=\pi/2$ with different twist angles $\alpha=\pi$ in Fig. \ref{fig3}(b) and $\alpha=3\pi/2$ in Fig. \ref{fig3}(c).
In Fig. \ref{fig3}(b), the pseudo-spin helical edge states remain gapless with $\alpha=\pi$.
Conversely, Fig. \ref{fig3}(c) demonstrates that the edge states are gapped by interlayer coupling with $\alpha=3\pi/2$.
Figures. \ref{fig3}(e,f) display the band structures of the twist bilayer CIs nanoribbon at the angle $\theta=3\pi/4$ with different twist angles $\alpha$.
Figure \ref{fig3}(e) shows gapped edge states with $\alpha=\pi$, whereas Fig. \ref{fig3}(f) reveals gapless edge states with $\alpha=3\pi/2$.
These results indicate that interlayer coupling does not destroy the gapless edge states when the condition $\theta=\alpha/2$ is satisfied.
Otherwise, the edge states are gapped by interlayer coupling,
resulting in a non-zero Dirac mass term.

\begin{figure}
	\centering
	\includegraphics[width=1\columnwidth,clip]{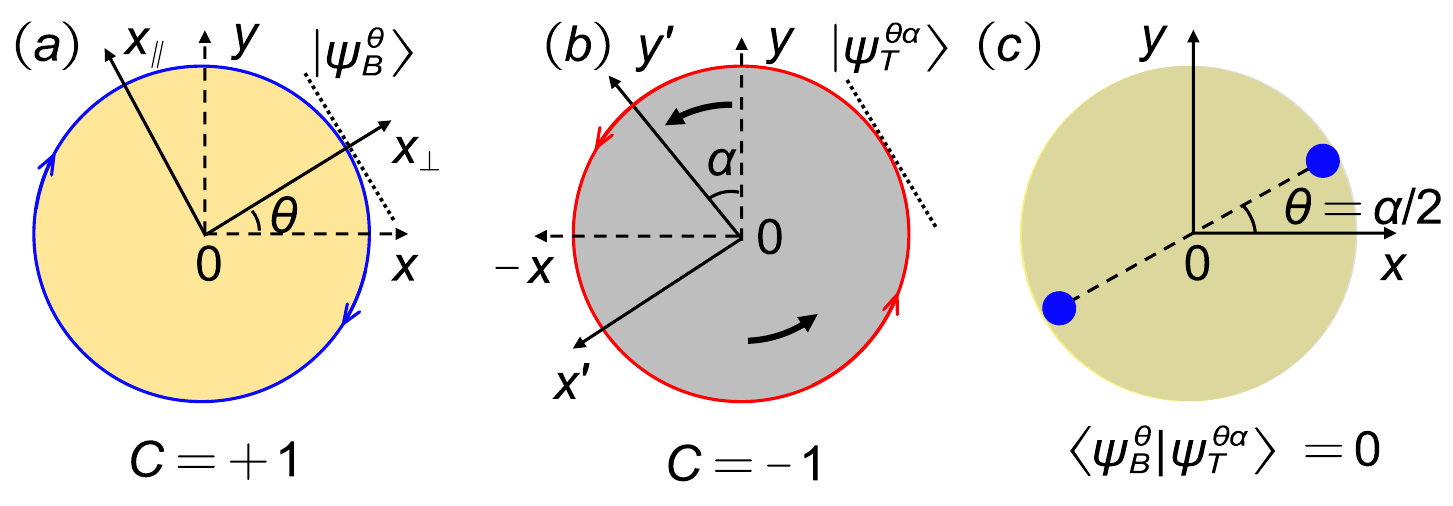}
	\caption{ (a) Schematic diagram of a vector normal and parallel to the tangential edge $\left| \psi_{B}^{\theta} \right>$ in a CI with $C=+1$. The angle $\theta$ defines the direction of the edge. (b) Schematic diagram of the twisted coordinate system and the tangential edge $\left| \psi_{T}^{\theta \alpha} \right>$ in a CI with $C=-1$. The twist angle is labeled as $\alpha$. (c) Schematic diagram of the coupled system appearing corner states at $\theta=\alpha/2$. Corner states occurrence satisfies the orthogonal condition $\left< \psi_{B}^{\theta}|\psi_{T}^{\theta \alpha} \right>=0$.
	}
	\label{fig4}
\end{figure}

To gain a thorough understanding of the second-order topological corner states, we study the chiral edge states in the
decoupled bilayer system. Combining Eqs. (\ref{eq2}) and (\ref{eq3}), the Hamiltonian for the top CI layer can be expressed as:
\begin{align}
  H_{T}(\mathbf{k})=&A(-k_{x} \cos {\alpha}-k_{y} \sin {\alpha}) \sigma_{x} \nonumber \\
  &+A(-k_{x} \sin {\alpha}+k_{y} \cos {\alpha}) \sigma_{y}
  +(M-B k^{2})\sigma_{z}.
  \label{eq5}
 \end{align}
We consider the same tangential edge at $\theta$ for both the top and bottom CI layers [see Figs. \ref{fig4}(a,b)].
For an arbitrary sample edge with normal direction $\mathbf{x}_{\bot}=(\cos\theta,\sin\theta)$, and assuming a half-infinite sample area $x_{\bot}<0$, the effective Hamiltonian can be written as \cite{Tan2022,Poata2023}:
\begin{align}
  H_{T}(\bf{k})=&A[\cos(\theta-\alpha)\sigma_{x}+\sin(\theta+\alpha)\sigma_{y}]k_{\bot}\nonumber \\
  &+A[\sin(\theta-\alpha)\sigma_{x}-\cos(\theta+\alpha)\sigma_{y}]k_{\parallel}\nonumber \\
  &+[M-B(k_{\bot}^{2}+k_{\parallel}^{2})]\sigma_{z}.
  \label{eq6}
 \end{align}
We replace $k_{\bot}$ with $-i\partial_{x_{\bot}}$ and decompose the effective Hamiltonian into two parts: $H_{T}(k)=H_{T}^{0}(k_{\bot})+H_{T}^{p}(k_{\parallel})$, where
\begin{align}
H_{T}^{0}(k_{\bot})=&-iA[\cos(\theta-\alpha)\sigma_{x}+\sin(\theta+\alpha)\sigma_{y}]\partial_{x_{\bot}}\nonumber \\
&+\left(M+B\partial_{x_{\bot}}^{2}\right)\sigma_{z},\nonumber \\
H_{T}^{p}(k_{\parallel})=&A[\sin(\theta-\alpha)\sigma_{x}-\cos(\theta+\alpha)\sigma_{y}]k_{\parallel}\nonumber \\
&-Bk_{\parallel}^{2}\sigma_{z}.
\label{eq7}
 \end{align}
The edge states can be obtained by solving the equation $H_{T}^{0}(k_{\bot})\psi_{T}(x_{\bot})=E\psi_{T}(x_{\bot})$ with the edge conditions $\psi_{T}(0)=\psi_{T}(-\infty)=0$. We assume a trial function of the form $\psi_{T}(x_{\bot})=e^{\lambda x_{\bot}} \xi_{T}$, where $\xi_{T}$ is a spinor.
The edge state of the top CI layer takes the form as
\begin{align}
\psi_{T}^{\theta \alpha}(x_{\bot})= \mathcal{N} \sin (\kappa_1 x_{\bot})e^{\kappa_{2} x_{\bot}}
e^{ik_{\parallel}x_{\parallel}}
\left[\begin{array}{c}
1\\
ie^{-i(\theta-\alpha)}
\end{array}\right]
\label{eq8}
 \end{align}
with normalization given by $\left|\mathcal{N} \right|^2=2 \left|\kappa_{2}(\kappa_{1}^{2}+\kappa_{2}^{2})/\kappa_{1}^{2}\right|$, where $\kappa_{1}=\sqrt{-(\frac{A}{2B})^2+\left|\frac{M}{B} \right|}$ and $\kappa_{2}=\frac{A}{2B}$.

Similarly, the eigenstates of the bottom CI layer at $\theta$ edge [see Fig. \ref{fig4}(a)] turn out to be:
\begin{align}
\psi_{B}^{\theta}(x_{\bot})= \mathcal{N} \sin (\kappa_1 x_{\bot})e^{\kappa_{2} x_{\bot}}
e^{ik_{\parallel}x_{\parallel}}
\left[\begin{array}{c}
1\\
-ie^{i\theta}
\end{array}\right]
\label{eq9}
 \end{align}
It is noteworthy that the occurrence of corner states corresponds to the orthogonal conditions of the edge states in the top and bottom layers $\left<\psi_{B}^{\theta}|\psi_{T}^{\theta \alpha}\right>=0$, which implies $\theta=\alpha/2, \alpha/2+\pi$, as shown in Fig. \ref{fig4}(c).
The edge states of the top and bottom CIs remain orthogonal only at edge $\theta=\alpha/2, \alpha/2+\pi$, which is the basis for the existence of corner states.

\begin{figure*}
	\centering
	\includegraphics[width=2.1\columnwidth,clip]{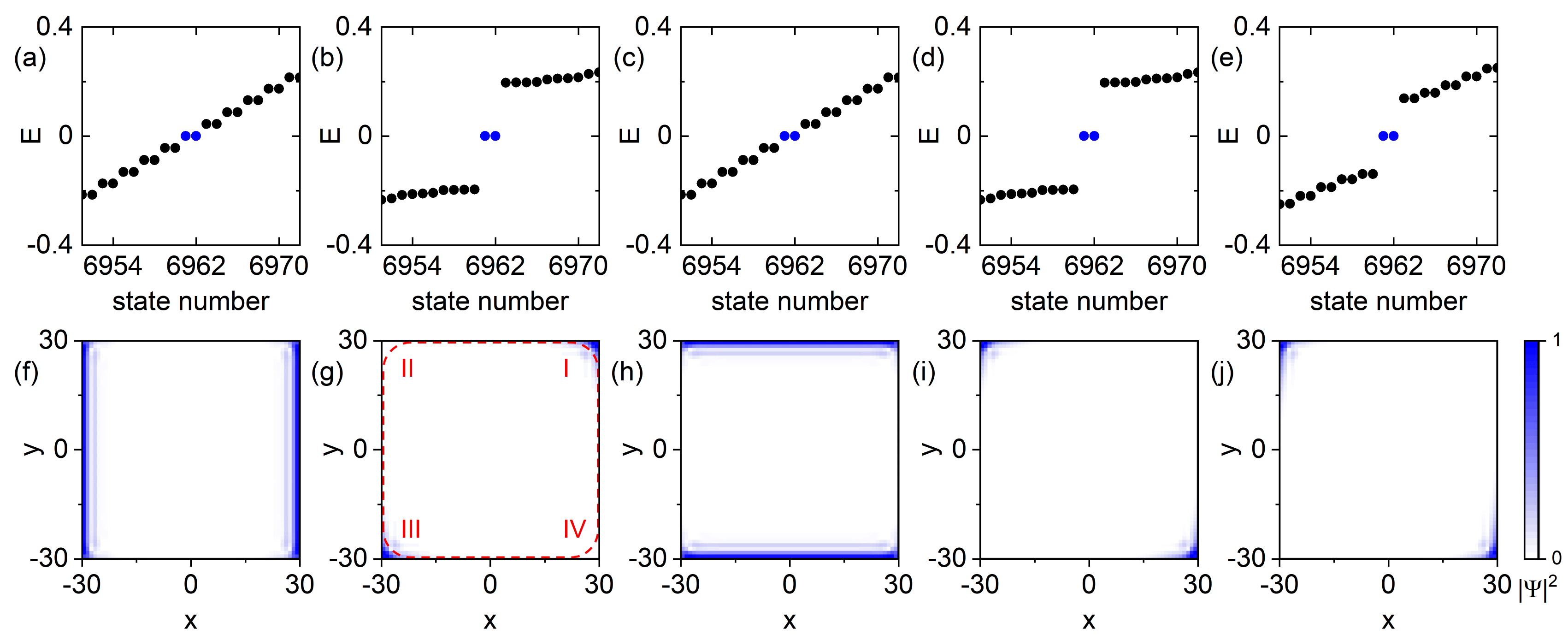}
	\caption{(a-e) Energy levels of square-shaped coupled bilayer CIs nanoflakes with different twist angles $\alpha$.
Blue dots correspond to zero-energy states.
(f-j) Probability distribution of the zero-energy states.
The brightness of blue is proportional to the square of the wave function [see color bar]. $I$--$IV$ in (g) denote the four corners.
The red dashed line in (g) displays that corners can be seen as a smooth evolution from one boundary to another.
The twist angle are $\alpha = 0$ for (a,f), $\alpha = \pi/2$ for (b,g), $\alpha = \pi$ for (c,h), $\alpha = 3\pi/2$ for (d,i), and $\alpha = 5\pi/3$ for (e,j). The nanoflake size is set to be $61 \times 61$ in (a-j). Other parameters are the same as those in Fig. \ref{fig2}(f).
}
	\label{fig5}
\end{figure*}

\subsection{\label{sec3C} Application of the universal theory}

To demonstrate the universality of our edge theory, we plot the energy levels and wave function distribution of square-shaped coupled nanoflakes with different twist angles $\alpha$ in Figs. \ref{fig5}.
The nanoflake size is set to be $61 \times 61$, and the twist angles are $\alpha = 0$ in Fig. \ref{fig5}(a,f), $\alpha = \pi/2$ in Fig. \ref{fig5}(b,g), $\alpha = \pi$ in Fig. \ref{fig5}(c,h), $\alpha = 3\pi/2$ in Fig. \ref{fig5}(d,i), and $\alpha = 5\pi/3$ in Fig. \ref{fig5}(e,j).
The blue dots in Figs. \ref{fig5}(a-e) correspond to zero energy states.
Figure \ref{fig5}(a) shows that the continuous states persist with the twist angle $\alpha=0$.
The wave function distribution reveals that these states extend at left and right edges with normal angles $\theta=\pi$ and $0$ [see Fig. \ref{fig5}(f)].
This is well consistent with the result in Sec. \ref{sec3B}
that the gapless edge states only exist at the edge with $\theta=\alpha/2$ and
$\theta=\alpha/2 +\pi$.

While $\alpha=\pi/2$, the two zero-energy in-gap states emerge and are almost distributed at the two corners $I$ and $III$, as shown in Figs. \ref{fig5}(b,g).
Below, we analyze why the zero-energy corner states appear
at the corners $I$ and $III$ and not at the corners $II$ and $IV$.
For a corner composed of two boundaries, we can consider it as a smooth evolution from one boundary to the other within a small range [see red dashed line in Fig. \ref{fig5}(g)].
Let us assume that the normal angles of the two boundaries of a corner are $\theta_1$ and $\theta_2$ with $\theta_1<\theta_2$.
According to our theory in Sec. \ref{sec3A} and Sec. \ref{sec3B},
the corner state appears at $\theta=\alpha/2$ and $\theta=\alpha/2+\pi$.
So when $\theta_1 < \frac{\alpha}{2} < \theta_2$ or
$\theta_1 < \frac{\alpha}{2}+\pi < \theta_2$,
there exists the corner state at this corner.
Conversely, if $\frac{\alpha}{2}$ and $\frac{\alpha}{2}+\pi$ are not in the
range of $[\theta_1,\theta_2]$, there is without the corner state at this corner.
The normal angles of the two boundaries for the corners $I$, $II$, $III$, and
$IV$ are $\theta^{I}_{1}=0$ and $\theta^{I}_{2}=\frac{\pi}{2}$,
$\theta^{II}_{1}=\frac{\pi}{2}$ and $\theta^{II}_{2}=\pi$,
$\theta^{III}_{1}=\pi$ and $\theta^{III}_{2}=\frac{3\pi}{2}$,
and $\theta^{IV}_{1}=\frac{3\pi}{2}$ and $\theta^{IV}_{2}=2\pi$, respectively
[see Fig. \ref{fig5}(g)].
For $\alpha=\pi/2$, the condition $\theta^{I}_{1}<\alpha/2<\theta^{I}_{2}$
($\theta^{III}_{1}<\alpha/2 +\pi <\theta^{III}_{2}$) satisfies,
leading to the appearance of the corner state at the corner $I$ ($III$).
The other two corners $II$ and $IV$ do not satisfy this relationship,
so there is no corner state at the corners $II$ and $IV$.

\begin{figure*}
	\centering
	\includegraphics[width=2.1\columnwidth,clip]{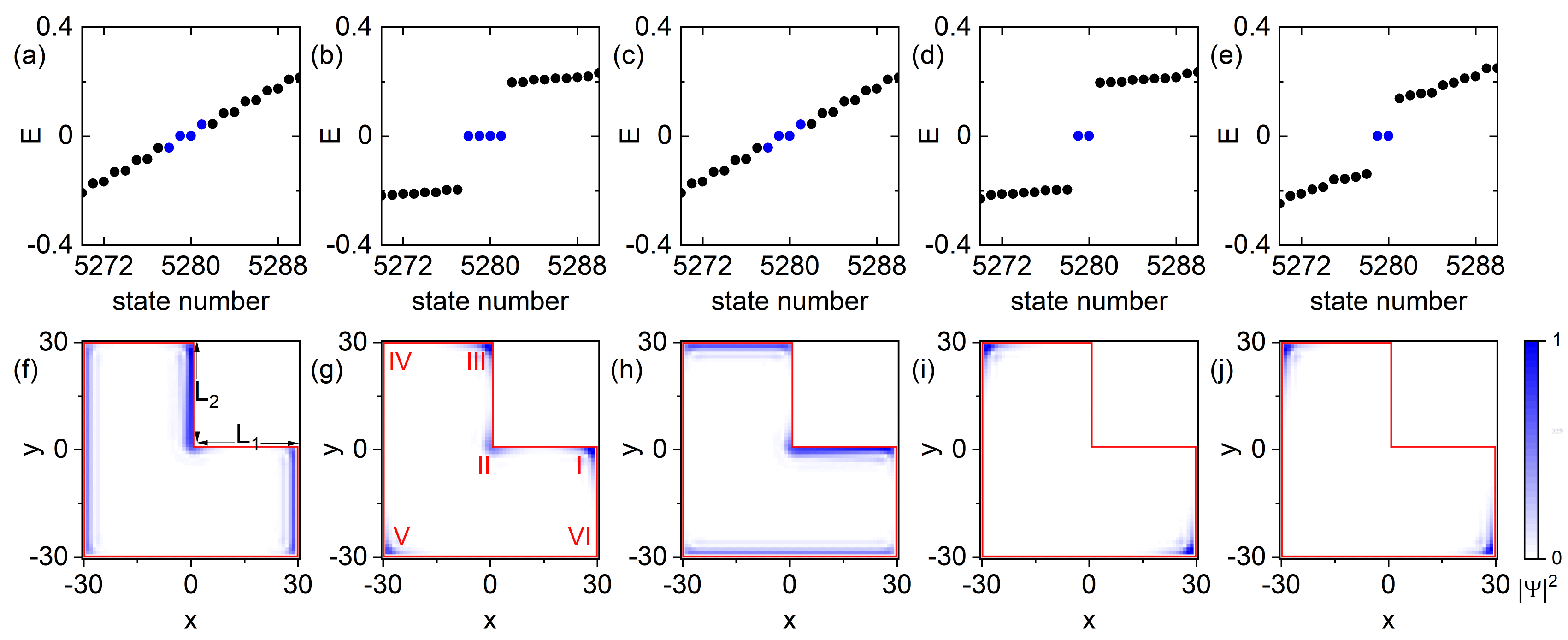}
	\caption{ (a-e) Energy levels of L-shaped coupled bilayer CIs nanoflakes with the different twist angles $\alpha$.
The L-shaped nanoflake is constructed from a $61\times 61$ square nanoflake with a hollow measuring $L_1 \times L_2=30\times 30$ at the upper right corner.
Blue dots correspond to near-zero energy states with the probability distribution shown in panels (f-j).
The brightness of blue is proportional to the square of the wave function [see color bar].
The red solid lines indicate the boundary of the L-shaped nanoflake,
and $I$--$VI$ in (g) indicate the six corners.
The twist angle are $\alpha = 0$ for (a,f), $\alpha = \pi/2$ for (b,g), $\alpha = \pi$ for (c,h), $\alpha = 3\pi/2$ for (d,i), and $\alpha = 5\pi/3$ for (e,j). Other parameters are selected the same as Fig. \ref{fig2}(f).
}
	\label{fig6}
\end{figure*}

While $\alpha=\pi$, as displayed in Fig. \ref{fig5}(c), the continuous states reappear in the coupled bilayer CIs system. But the edge states along the $y$ direction are gapped, while those along the $x$ direction with $\theta=\pi/2$ and $\theta=3\pi/2$ remain gapless [see Fig. \ref{fig5}(h)], which is well consistent with our theory.
For $\alpha=3\pi/2$, two zero-energy in-gap states are observed,
as displayed by blue dots in Fig. \ref{fig5}(d).
The normal angles of the boundaries where the gapless states are located are $\theta=\alpha/2=3\pi/4$ and $\theta=\alpha/2+\pi=7\pi/4$.
Corners $II$ and $IV$ satisfy the conditions for the existence of the corner states, and Fig. \ref{fig5}(i) visualizes that the in-gap state appears at corners $II$ and $IV$.
In Fig. \ref{fig5}(e), one can see that the two zero-energy in-gap states still appear for a more general twist angle $\alpha=5\pi/3$, and the positions of in-gap states remain at corners $II$ and $IV$, as shown in Fig. \ref{fig5}(j).
These results demonstrate that the locations of corner states can be accurately predicted by our theory.
If the normal angles ($\theta_1$, $\theta_2$) of the two sides of the corner satisfy the relation $\min(\theta_{1},\theta_2) <\alpha/2 (\alpha/2 +\pi)<\max(\theta_{1},\theta_2)$, there is a zero-dimensional corner state at this corner.
This approach can explain all previous works on interlayer coupling induced second-order corner states \cite{Mandal2024,Liu2024,Liu2024a}, demonstrating the universality of this theory.

To achieve multiple corner states, we design an L-shaped nanoflake with a lattice hollow of size $L_1 \times L_2=30*30$ within a $61*61$ square nanoflake, as shown in Fig. \ref{fig6}(f-j).
In Figs. \ref{fig6}(a-e), we plot the energy levels of the L-shaped coupled bilayer nanoflakes with different twist angles $\alpha$.
The wave function distributions of the corresponding near zero energy states are shown in turn by Fig. \ref{fig6}(f-j).
The twist angle are set as $\alpha = 0$ in Fig. \ref{fig6}(a,f), $\alpha = \pi/2$ in Fig. \ref{fig6}(b,g), $\alpha = \pi$ in Fig. \ref{fig6}(c,h), $\alpha = 3\pi/2$ in Fig. \ref{fig6}(d,i), and $\alpha = 5\pi/3$ in Fig. \ref{fig6}(e,j).
There are continuous states in the L-shaped nanoflake while $\alpha=0$ [see Fig. \ref{fig6}(a)].
It can be seen from Fig. \ref{fig6}(f) that the continuous states are almost extended along the $y$ axis, which is well consistent with our theory.
As displayed in Fig. \ref{fig6}(b), there arise four zero-energy in-gap states (highlighted in blue dots) for $\alpha=\pi/2$.
By analyzing the local density of states of the zero-energy states in Fig. \ref{fig6}(g), one can see that these states are localized at four corners $I$, $II$, $III$ and $V$.
In fact, there are six corners marked as $I$--$VI$ in the L-shaped nanoflake [see Fig. \ref{fig6}(g)]. The normal angles of two boundaries forming corners $I$--$VI$ are respectively
$\theta^{I}_1=0$ and $\theta^{I}_2=\frac{\pi}{2}$,
$\theta^{II}_1=\frac{\pi}{2}$ and $\theta^{II}_2=0$,
$\theta^{III}_1=0$ and $\theta^{III}_2=\frac{\pi}{2}$,
$\theta^{IV}_1=\frac{\pi}{2}$ and $\theta^{IV}_2=\pi$,
$\theta^{V}_1=\pi$ and $\theta^{V}_2=\frac{3\pi}{2}$,
and $\theta^{VI}_1=\frac{3\pi}{2}$ and $\theta^{VI}_2=2\pi$.
For the corners $I$, $II$, and $III$,
they are satisfied $\min(\theta^{I/II/III}_{1},\theta^{I/II/III}_{2})
< \frac{\alpha}{2}  < \max(\theta^{I/II/III}_{1},\theta^{I/II/III}_{2})$,
leading to the zero-energy states being bound to be here.
The corner $V$ also bounds the zero-energy in-gap states because of
$\theta^{V}_{1} < \frac{\alpha}{2} +\pi < \theta^{V}_{2}$.
For the corners $IV$ and $VI$, $\frac{\alpha}{2}$ and $\frac{\alpha}{2}+\pi$
are not in the range $[\theta^{IV/VI}_1,\theta^{IV/VI}_2]$,
so there is without corner state at the corners $IV$ and $VI$.
As the twist angle turns to be $\alpha=\pi$,
the continuous states appear again, but they are distributed into the three $x$-direction edges of the L-shaped nanoflake [see Figs. \ref{fig6}(c,h)].
As the twist angle increases to $\alpha=3\pi/2$, two zero energy in-gap states appear, as displayed by the blue dots in Fig. \ref{fig6}(d).
At $\alpha=3\pi/2$, the normal angle of the boundaries where the gapless states are located is changed to be $\theta=3\pi/4, 7\pi/4$.
Corners $II$ and $IV$ satisfy the conditions for the existence of the corner states, and Figure \ref{fig6}(i) visualizes that the in-gap state appears at corners $II$ and $IV$.
For $\alpha=5\pi/3$, two zero-energy in-gap states remain, primarily located at corners $IV$ and $VI$ [see Figs.~\ref{fig6}(e,j)].
The above results demonstrate that a controllable number of second-order corner states can be induced at arbitrary positions by adjusting the boundary conditions and twist angle.
Hence, our findings provide a universal theory for inducing second-order topological corner states.

\section{\label{sec4}summary}

In summary, we propose an analytic approach to understanding second-order topological corner states induced by interlayer coupling.
This approach involves a coupled system of top and bottom nanoflakes with opposite Chern numbers, which can be realized through spatial inversion and twisting operations.
Our findings reveal that for a fixed twist angle $\alpha$, only the edge with its normal angle $\theta$ at $\alpha/2$ and $\alpha/2+\pi$ remain gapless, while the rest of the edges are gapped.
So second-order topological corner states emerge at a specific edge with its normal angle $\theta=\alpha/2, \alpha/2+\pi$.
In a finite-size nanoflake, if the normal angles ($\theta_1$, $\theta_2$) of the two sides of the corner satisfy the relation $\min(\theta_{1},\theta_2) <\alpha/2 (\alpha/2 +\pi)<\max(\theta_{1},\theta_2)$, there is a zero-dimensional corner state at this corner.
So the positions of the corner states can be precisely controlled by varying the twist angle $\alpha$.
This universal theory not only clarifies the mechanism behind previously studied interlayer coupling to achieve topological corner states but also applies to various shapes.

\begin{acknowledgments}
This work was financially supported by the National Natural Science Foundation of China (Grants No. 12374034, No. 11921005, and No. 12074097),
Natural Science Foundation of Hebei Province (Grant No. A2024205025),
the Innovation Program for Quantum Science and Technology (Grant No. 2021ZD0302403), and the Strategic Priority Research Program of Chinese Academy of Sciences (Grant No. XDB28000000).
We also acknowledge the High-performance Computing Platform of Peking University for providing computational resources.
\end{acknowledgments}


\end{document}